\documentclass{PoS}

\usepackage{subfig}
\newcommand{\subfigure}{\subfloat}
\newcommand{\Subref}[1]{\protect\subref{#1}}
\usepackage{url}
  
\title{Instrumentation of the Very Forward Region at Future Linear Colliders -- design and R\&D by the FCAL Collaboration}
 
\ShortTitle{Instrumentation of the Very Forward Region at Future Linear Colliders}

\author{\speaker{Iftach Sadeh}%
         \thanks{On behalf of the FCAL Collaboration~\cite{bib1}.}\\
         School of Physics and Astronomy, The Raymond and Beverly Sackler Faculty of Exact Sciences, Tel Aviv University, Tel Aviv, Israel.\\
        E-mail: \email{sadeh@alzt.tau.ac.il}}

\abstract{
Two finely segmented and very compact  calorimeters are foreseen for the instrumentation of the very forward region of the ILC
and the CLIC detectors;
a luminometer to measure the rate of low angle Bhabha scattering events with
a precision better than~$10^{-3}$ and, at lower polar angles, a detector for fast luminosity estimate.
Both devices will improve the hermeticity of the detector in searches for new particles.
Due to the expected high occupancy, fast front-end electronics is needed.
The calorimeter closest to the beampipe  will be
hit by a large amount of beamstrahlung remnants, so that radiation hard sensors are required. 
Results on the performance of the sensor and readout electronics prototypes are summarized.
}

\FullConference{35th International Conference of High Energy Physics\\
		 July 22-28, 2010\\
		 Paris, France}

\begin{document}

\section{Introduction}
 
Two calorimeters are foreseen in the very forward region 
for the two proposed $e^+e^-$ linear colliders, the ILC and
CLIC\footnote{ This
report summarizes R\&D work performed by the FCAL collaboration~\cite{bib1}. Please see~\cite{bib2} for
further details and references.};
LumiCal will be used to determine the luminosity by measuring the rates of Bhabha scattering events,
and BeamCal will provide a fast estimate of the 
luminosity and control of beam parameters by measuring beamstrahlung radiation. 
Both are designed as cylindrical sensor-tungsten sandwich
electromagnetic (EM) calorimeters, each consisting of 30~absorber disks of 3.5~mm thickness
interspersed with sensors.
Sensor layers are segmented radially and azimuthally 
into pads, and front-end (FE) ASICs are positioned at the outer radii.

In order to reach the required accuracy for measurement of the luminosity
(better than~$10^{-3}$ for most ILC physics channels and $\sim10^{-4}$ 
for the GigaZ program), LumiCal sensors must be finely granulated.
Regarding BeamCal, the main challenge is the development of radiation hard sensors, due to the high
radiation dose expected at low polar angles.
In addition, all detectors in the very forward region have to tackle relatively high 
occupancy, requiring special FE electronics.

\section{Sensor Development and Front-End Electronics Design}

\begin{figure}[htp]
\centering
\subfigure[]{\label{fig:sensors1}\includegraphics[height=.17\textheight,width=.3\textwidth]{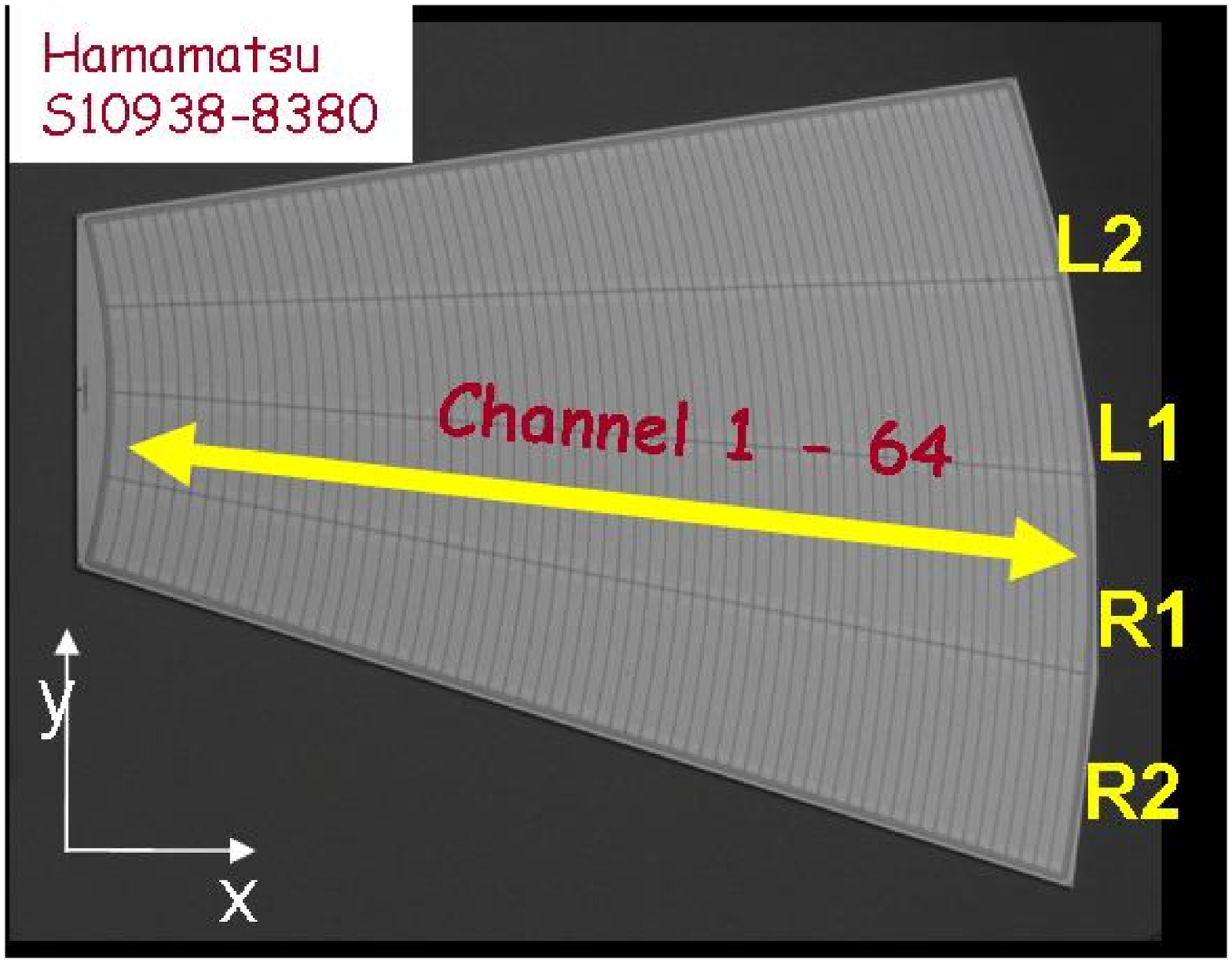}} \hspace{2mm}
\subfigure[]{\label{fig:sensors2}\includegraphics[height=.3\textwidth,width=.17\textheight,angle=90]{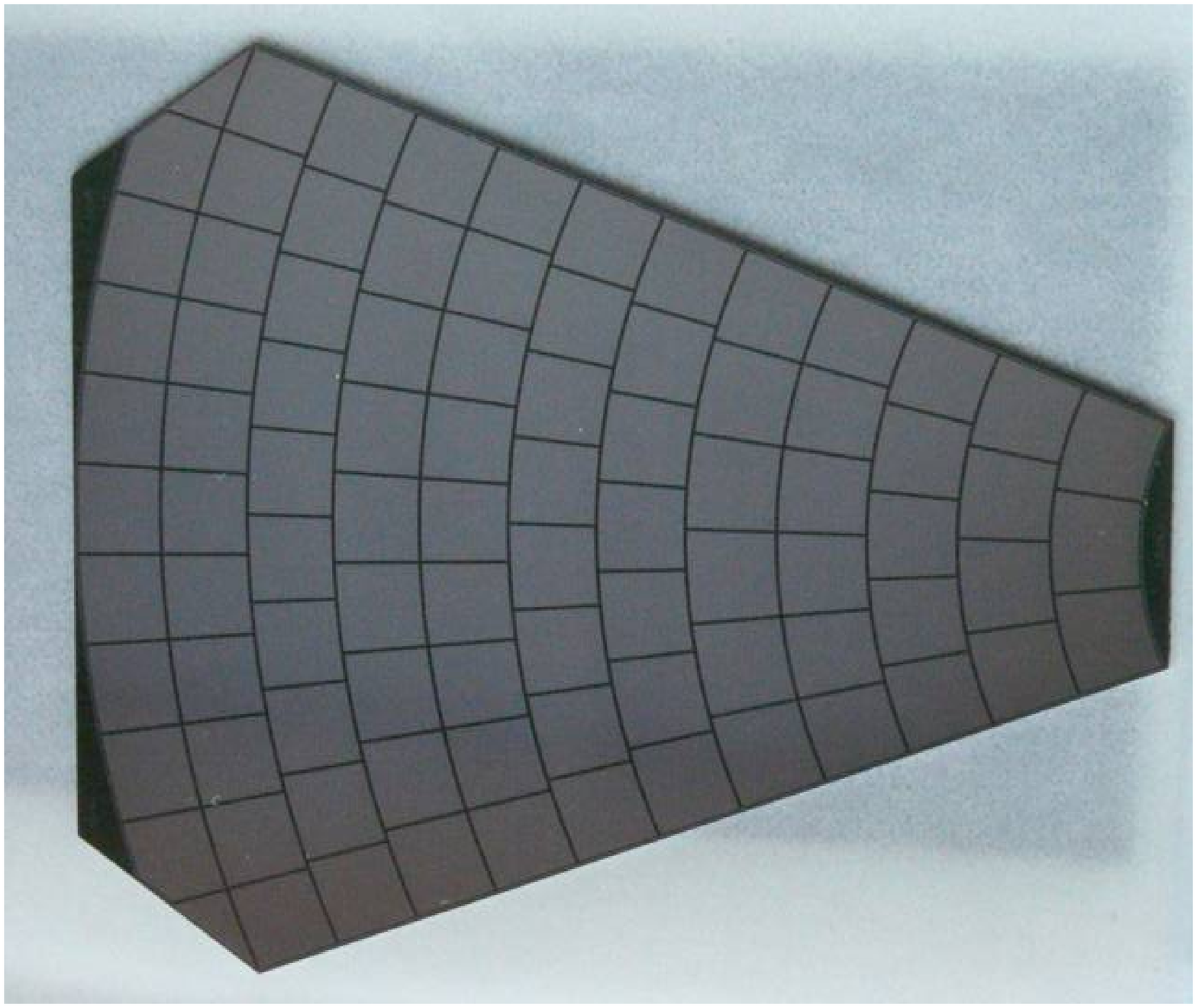}}\hspace{1mm}
\subfigure[]{\label{fig:sensors3}\includegraphics[trim=0mm 0mm 0mm 10mm,clip,height=.17\textheight,width=.3\textwidth]{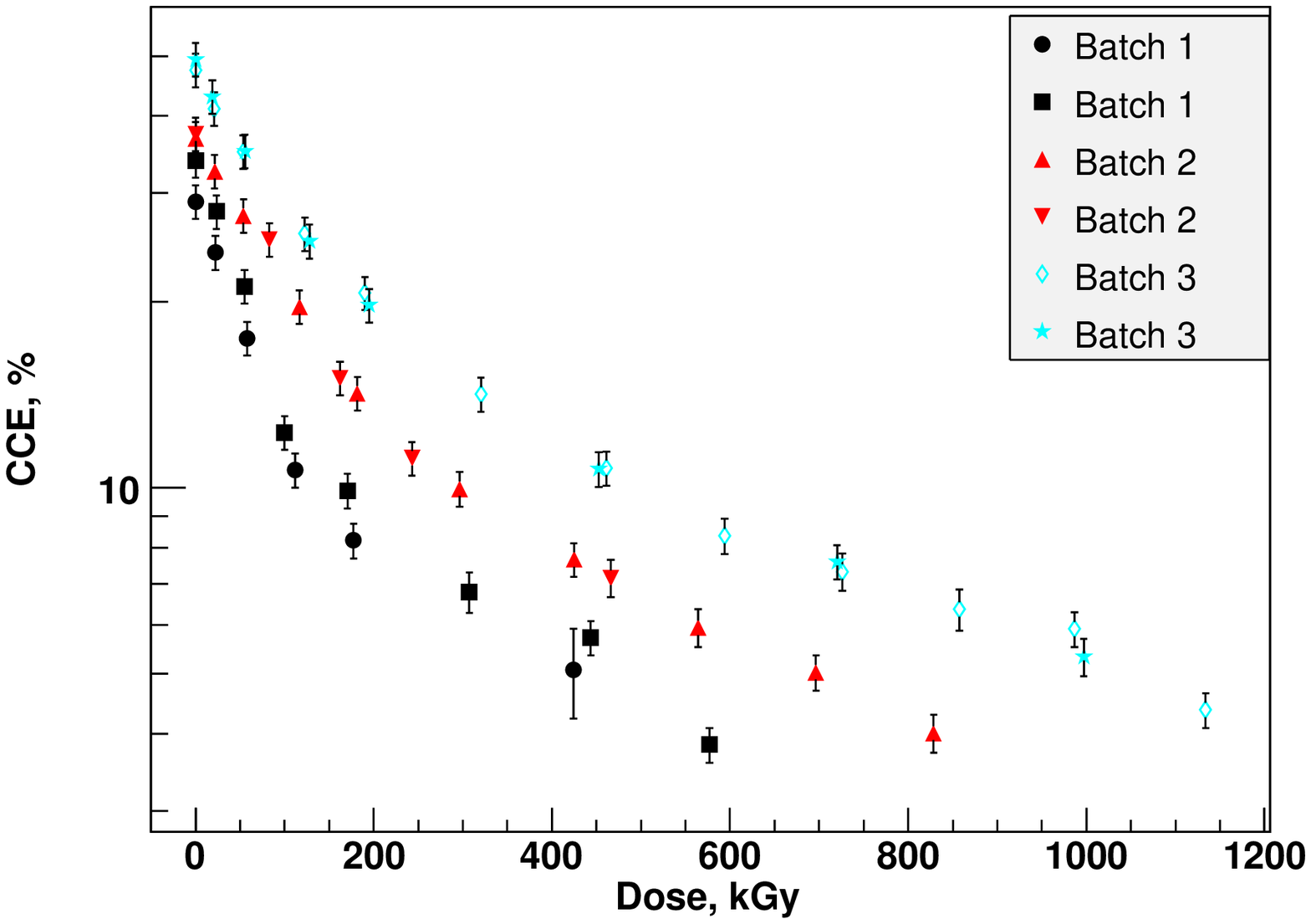}} 
\caption{Prototype silicon sensors for LumiCal \Subref{fig:sensors1} and GaAs sensors for BeamCal \Subref{fig:sensors2};
the CCE as a function of the absorbed dose for the GaAs sensors with different donor concentrations \Subref{fig:sensors3}.
The donor is tellur for batches~1 and~2 and tin for batch~3.}
\label{fig:sensors}
\end{figure}

Prototypes of LumiCal sensors, shaped as ring segments
of 30$^\circ$, have been designed 
and then manufactured by Hamamatsu Photonics (see Fig.~\ref{fig:sensors1}).
The thickness of the n-type silicon bulk is 320~$\mu$m. The pitch 
of the concentric p$^+$ pads is 1.8~mm and the gap between two pads is 0.1~mm.

So far polycrystalline CVD~diamond sensors of 
1~cm$^2$ size and larger sectors of GaAs pad sensors (see Fig.~\ref{fig:sensors2}) have been studied for BeamCal.
Since large area CVD~diamond sensors are extremely expensive, 
they may be used only at the innermost part of BeamCal.
The charge collection efficiency, CCE, was measured for several GaAs batches
with different concentrations of dopants, which were irradiated by up to 1.5~MGy (see Fig.~\ref{fig:sensors3}).
The leakage current of a pad at room temperature before irradiation 
was about 500~nA at an applied voltage of 100~V. After the irradiation the leakage current
approached 1~$\mu$A.
For polycrystalline diamond sensor samples the linearity of the response, the
leakage current and the signal collection efficiency have been investigated
as well. 
The signal size depends linearly on the number of charged particles crossing the sensors
for up to 5$\times$10$^6$ particles in 10~ns.
The leakage current at room temperature depends only slightly on the absorbed dose up to 7~MGy.
The CCE rises by a factor of two for doses between 0.5 to 1~MGy, then 
drops  smoothly approaching the respective value of a non-irradiated sensor. 

\begin{figure}[htp]
\centering
\subfigure[]{\label{fig:fe1}\includegraphics[height=.17\textheight,width=.45\textwidth]{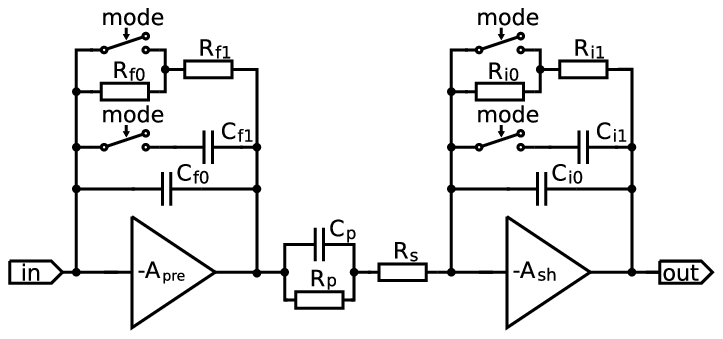}}
\subfigure[]{\label{fig:fe2}\includegraphics[height=.17\textheight,width=.45\textwidth]{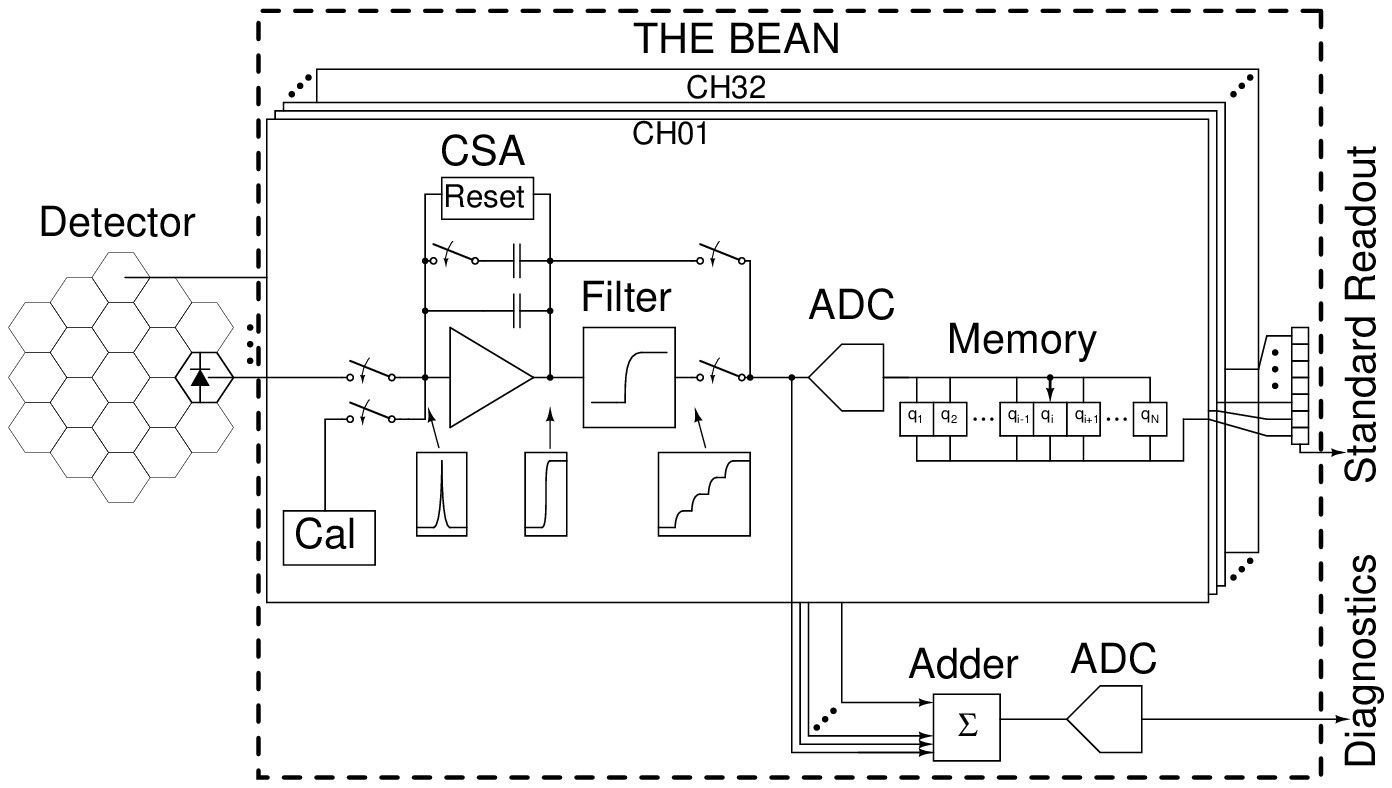}}
\caption{
Block diagram of a single FE channel for LumiCal \Subref{fig:fe1} and for BeamCal \Subref{fig:fe2}.
}
\label{fig:fe}
\end{figure}

The FE ASIC of LumiCal
is designed to work in two modes, the physics mode and the calibration mode. 
In the physics mode, EM showers will be measured with large energy depositions on the pads, processing 
signals of up to 6~pC per channel. In the calibration mode, signals from minimum ionizing particles
will be measured, with signals above 2~fC. The signal will be reconstructed using at least 8~bit precision.
The FE architecture comprises a charge sensitive amplifier, 
a pole-zero cancellation circuit and a shaper, as shown in Fig.~\ref{fig:fe1}.
The LumiCal ADC was designed using a 1.5-bit per stage pipeline architecture with power switching.
Test results of the ADC linearity for the prototype show good linearity,
and a signal-to-noise ratio of about 58~dB is obtained in the frequency range up to almost 25~MHz.

The BeamCal ASIC, designed for 180~nm TSMC technology, will be able to handle 
32~channels. The two modes of operation require a FE circuit capable of a wide performance 
envelope; high slew rate for standard data taking, and low noise for calibration. 
A digital memory array will be used to store the data from 
all collisions in each bunch train, requiring a sampling rate of 3.25~MHz per 
channel. This is achieved by 10-bits successive approximation analog-to-digital converters,
using one converter per channel.
The filter for calibration operation has been implemented using switched-capacitor 
circuits.
In standard data taking operation, an adequate noise power is effectively achieved 
by means of a slow reset-release technique.
Test results of the ADC prototype show good linearity and are consistent 
with unit capacitance, matching better than 0.1\%.

\bibliographystyle{ieeetr}  \bibliography{bib}   

\begin{thebibliography}{1}

\bibitem{bib1}
{The FCAL Collaboration} {URL:} \url{http://www-zeuthen.desy.de/ILC/fcal/}.

\bibitem{bib2}
H.~{A}bramowicz \textit{et al.}, ``{Forward Instrumentation for ILC
  Detectors},'' {\em ArXiv e-prints}, Sept. 2010.
\newblock {URL:} \url{http://arxiv.org/abs/1009.2433}, {Submitted for
  publication in JINST}.

\end{thebibliography}

%
%


\end{document}